% =======================LETRAS HUECAS============================
\newfam\msbfam
\font\twlmsb=msbm10 at 12pt
\font\eightmsb=msbm10 at 8pt
\font\sixmsb=msbm10 at 6pt
\textfont\msbfam=\twlmsb
\scriptfont\msbfam=\eightmsb
\scriptscriptfont\msbfam=\sixmsb
\def\cj{\fam\msbfam}

\def\C{{\cj C}}

\def\R{{\cj R}}

% =======================LETRAS HUECAS============================
\newfam\msbfam
\font\twlmsb=msbm10 at 12pt
\font\eightmsb=msbm10 at 8pt
\font\sixmsb=msbm10 at 6pt
\textfont\msbfam=\twlmsb
\scriptfont\msbfam=\eightmsb
\scriptscriptfont\msbfam=\sixmsb
\def\cj{\fam\msbfam}

\def\C{{\cj C}}

\def\R{{\cj R}}

\centerline{\bf Extended Newman-Janis algorithm for rotating and Kerr-}

\

\centerline{\bf Newman de Sitter and anti de Sitter metrics} 

\

\centerline{E. J. Gonzalez de Urreta* and M. Socolovsky**}

\

\centerline{\it  Instituto de Ciencias, Universidad Nacional de General Sarmiento}
\centerline{\it Juan Mar\'\i a Guti\'errez 1150 (B1613GSX), Los Polvorines, Pcia. de Buenos Aires, Argentina} 

\

{\bf Abstract.} {\it The Newman-Janis algorithm is well known to provide rotating black holes solutions to Einstein's equations from static seeds, through a complexification of a radial and a time coordinates. However, an ambiguity remains for the replacement of the $r^{-1}$ and $r^{-2}$ powers of the radial coordinate. We show here that the two cases are unified by a simple expression which allows its extension to the $r^{2}$ power, characteristic of the de Sitter ($dS$) and anti de Sitter ($AdS$) spacetimes. The formula leads almost automatically to the Kerr and Kerr-Newman-$dS$ and -$AdS$ metrics.}

\


\

{\bf 1. Introduction}

\

A usual criticism to the Newman-Janis algorithm (NJA) [1] for generating rotating metrics from seed static ones, is its apparent arbitrariness [2] in the replacement of powers or products of the complexified radial coordinate $r$ and its complex conjugate $\bar{r}$. So, ${{1}\over{r}}$ is replaced by ${{1}\over{2}}({{1}\over{r}}+{{1}\over{\bar{r}}})$ and ${{1}\over{r^2}}$ by ${{1}\over{r\bar{r}}}={{1}\over{\vert r\vert^2}}$. The de Sitter ($dS$) [3] and anti de Sitter ($AdS$) [4] metrics have an $r^2$ in the numerator (see eqs.(7) and (18)) and it is not clear, in this case, which would be the correct replacement.

\

It is easy to see that after the complexification $$\R\ni r\longrightarrow \C\ni r=r^\prime-ibcos\theta^\prime \eqno{(1)}$$ with $r^\prime\in(-\infty,+\infty)$, $\theta^\prime=\theta\in[0,\pi]$, and $b=const.>0$ to be interpreted as the rotation parameter (angular momentum/unit gravitational mass in the Kerr-de Sitter ($KdS$), Kerr-anti de Sitter ($KAdS$), Kerr-Newman de Sitter ($KNdS$) and Kerr-Newman anti de Sitter ($KNAdS$) cases), $$r^p\longrightarrow {{(Re(r))^{p+2}}\over{\vert r\vert^2}}={{{r^\prime}^{p+2}}\over{{r^\prime}^2+b^2cos^2\theta^\prime}}\eqno{(2)}$$ reproduces, for $p=-1,-2$, the above replacements: $$r^{-1}\longrightarrow {{Re(r)}\over{\vert r\vert^2}}={{r^\prime}\over{\vert r\vert^2}}={{1}\over{2}}({{1}\over{r}}+{{1}\over{\bar{r}}}),\eqno{(3)}$$ $$r^{-2}\longrightarrow {{1}\over{\vert r\vert^2}}={{1}\over{r\bar{r}}}.\eqno{(4)}$$ This suggests the extension to the case $p=2$: $$r^2\longrightarrow {{(Re(r))^4}\over{\vert r\vert^2}}={{{r^\prime}^4}\over{\vert r\vert^2}},\eqno{(5)}$$ which, as we will see, leads to the metrics for $rdS$ (rotating de Sitter) [5] and $rAdS$ (rotating anti de Sitter), and also for $KdS$, $KAdS$, $KNdS$, and $KNAdS$ [6]. 

\

In what follows all spacetimes are 4-dimensional, and all quantities are expressed in geometrical units.

\

{\bf 2. Schwarzschild anti de Sitter metric ($SAdS$)}

\

The $SAdS$ metric is given by $$ds^2_{SAdS}=f_{SAdS}dt^2-f_{SAdS}^{-1}dr^2-r^2d\Omega_2^2\eqno{(6)}$$ with $$f_{SAdS}=1-{{2M}\over{r}}+{{r^2}\over{a^2}},\eqno{(7)}$$ where $r>0$, $t\in (-\infty,+\infty)$, $d\Omega^2_2=d\theta^2+sin^2\theta d\varphi^2$, $\varphi\in[0,2\pi)$, $M$ is the mass of the Schwarzschild black hole and $a$ is the curvature radius of the $AdS$ space, corresponding to an atractive cosmological constant $$\Lambda_{AdS}=-{{3}\over{a^2}}.\eqno{(8)}$$ $f_{SAdS}(r)$ has no extrema for positive $r$, while $$f_{SAdS}\longrightarrow \{ \matrix{+\infty, & r\to+\infty\cr -\infty, & r\to 0_+}.$$ Then, it has a unique zero which is the position of the horizon: $$r_h(M,a)=(Ma^2)^{1/3}((1+\sqrt{1+{{a^2}\over{27M^2}}})^{1/3}+(1-\sqrt{1+{{a^2}\over{27M^2}}})^{1/3}), \ \ f_{SAdS}(r_h)=0.\eqno{(9)}$$ An expansion in ${{M}\over{a}}$ (typically $\ll$ 1) gives the deviation of $r_h$ from the Schwarzschild value $2M$: $$r_h=2M(1-\sqrt{3}{{M}\over{a}})\to 2M \ \ as \ \ a\to +\infty. \eqno{(10)}$$ The surface gravity $\kappa_{SAdS}$ at $r_h$ can be obtained from the calculation of the 4-accelerations of static observers, or through the use of the Rindler approximation in the neighborhood of the horizon, $r=r_h+{{\alpha}\over{r_h}}\rho^2$ [7], with $\alpha\in\R$, and neglecting terms of $O(\rho^4)$; the result is $$\kappa_{SAdS}={{M}\over{r_h^2}}+{{r_h}\over{a^2}}\to {{1}\over{4M}}=\kappa_S \ \ as \ \ a\to +\infty.\eqno{(11)}$$ With the choice $\alpha={{1}\over{2}}\kappa_{SAdS}r_h$, the time-radial part of the metric is the Rindler metric: $$ds^2_{SAdS}(\rho)\vert_{time-radial}=(\kappa_{SAdS}\rho)^2dt^2-d\rho^2.\eqno{(12)}$$ The Hawking temperature at $r_h$ is given by $$T_h={{\kappa_{SAdS}}\over{2\pi}}.\eqno{(13)}$$ (For the global embedding Minkowskian spacetime (GEMS) approach to this calculation, see ref. [8].)

\

In Eddington-Finkelstein retarded coordinates $(u,r,\theta,\varphi)$ [9], with $$dt=du+{{dr}\over{f_{SAdS}}},\eqno{(14)}$$ $u\in(-\infty,+\infty)$ and $r,\theta,\varphi$ as before, the $SAdS$ metric is $$ds^2_{SAdS}=f_{SAdS}du^2+2dudr-r^2d\Omega_2^2.\eqno{(15)}$$ The anti de Sitter metric is obtained setting $M=0$ i.e. with $$f_{AdS}=1+{{r^2}\over{a^2}}.\eqno{(16)}$$

\

{\bf 3. Schwarzschild de Sitter metric ($SdS$)}

\

The $SdS$ metric is given by $$ds^2_{SdS}=f_{SdS}dt^2-f_{SdS}^{-1}dr^2-r^2d\Omega_2^2\eqno{(17)}$$ with $$f_{SdS}(r)=1-{{2M}\over{r}}-{{r^2}\over{a^2}},\eqno{(18)}$$ where now $a$ is the curvature radius of the $dS$ space corresponding to a repulsive cosmological constant $$\Lambda_{dS}=+{{3}\over{a^2}}.\eqno{(19)}$$ Depending on the relation between $M$ and $a$ the $SdS$ metric has no horizon, one horizon, or two horizons. We shall discuss the latest case: $r_-$: black hole horizon, and $r_+$: cosmological horizon, which occur for $${{M}\over{a}}<{{1}\over{3\sqrt{3}}}\Longleftrightarrow M\sqrt{\Lambda_{dS}}<{{1}\over{3}}.\eqno{(20)}$$ $r_\pm$ are given by the two positive real zeros of (18): $$r_-={{2a}\over{\sqrt{3}}}cos(\varphi_0+{{4\pi}\over{3}}),\eqno{(21)}$$ $$r_+={{2a}\over{\sqrt{3}}}cos(\varphi_0),\eqno{(22)}$$ with $$\varphi_0={{1}\over{3}}arccos({{-3\sqrt{3}M}\over{a}})\in({{\pi}\over{6}},{{\pi}\over{3}}).\eqno{(23)}$$ Clearly, $r_-<r_+$. At $r_0=(Ma^2)^{1/3}$, with $r_-<r_0<r_+$, $f_{SdS}$ has a relative maximum $$f_{SdS}(r_0)=1-3({{M}\over{a}})^{2/3}>0.\eqno{(24)}$$ (An absolute maximum is $+\infty$, but occurs for $r\to 0_-$.)

\

The Rindler approximations outside but close to the black hole horizon: $r=r_-+{{-{\kappa_{SdS}}_-}\over{2}}\rho^2+O(\rho^4)$, and inside but close to the cosmological horizon $r=r_++{{-{\kappa_{SdS}}_+}\over{2}}\rho^2+O(\rho^4)$, allow us to compute the surface gravities: $${\kappa_{SdS}}_-=-{{M}\over{r_-^2}}+{{r_-}\over{a^2}}<0,\eqno{(25)}$$ $${\kappa_{SdS}}_+=-{{M}\over{r_+^2}}+{{r_+}\over{a^2}}>0,\eqno{(26)}$$ with $${ds^2_{SdS}(\rho)\vert_{time-radial}}_\pm=({\kappa_{SdS}}_\pm\rho)^2dt^2-d\rho^2.\eqno{(27)}$$ The Hawking temperatures at $r_\pm$ are given by $$T_\pm={{\vert{\kappa_{SdS}}_\pm\vert}\over{2\pi}}.\eqno{(28)}$$ In Eddington-Finkelstein retarded coordinates, $$ds^2_{SdS}=f_{SdS}du^2+2dudr-r^2d\Omega_2^2.\eqno{(29)}$$ The de Sitter metric is obtained from (29) setting $M=0$, i.e. replacing $f_{SdS}$ by $$f_{dS}=1-{{r^2}\over{a^2}}.\eqno{(30)}$$

\

{\bf 4. Rotating - de Sitter ($rdS$) and - anti de Sitter ($rAdS$) metrics}

\

We can unify the treatments of both metrics if we denote $ds^2_{dS}$ and $ds^2_{AdS}$ by $$ds^2_\Lambda=f_\Lambda du^2+2dudr-r^2d\Omega_2^2\eqno{(31)}$$ where $$f_\Lambda=1-{{\Lambda r^2}\over{3}}\eqno{(32)}$$ with $\Lambda=\Lambda_{dS}$ given by (19) and $\Lambda=\Lambda_{AdS}$ given by (8). The metric corresponding to (31) is given by $${g_{\mu\nu}}_\Lambda=\pmatrix{f_\Lambda & 1 & 0 & 0 \cr 1 & 0 & 0 & 0 \cr 0 & 0 & -r^2 & 0 \cr 0 & 0 & 0 & -r^2sin^2\theta} \eqno{(33)}$$ with inverse $$g^{\mu\nu}_\Lambda=\pmatrix{0 & 1 & 0 & 0 \cr 1 & -f_\Lambda & 0 & 0 \cr 0 & 0 & -r^{-2} & 0 \cr 0 & 0 & 0 & -r^{-2}sin^{-2}\theta}.\eqno{(34)}$$ It is easily verified that the inverse metric defined by $$\tilde{g}^{\mu\nu}=(l^\mu n^\nu+l^\nu n^\mu)-(m^\mu\bar{m}^\nu+m^\nu\bar{m}^\mu),\eqno{(35)}$$ where $(l,n,m,\bar{m})$ is the null tetrad given by $$l^\mu=(0,1,0,0), \ n^\mu=(1,-{{f_\Lambda}\over{2}},0,0), \ m^\mu={{1}\over{\sqrt{2}r}}(0,0,1,{{i}\over{sin\theta}}), \ \bar{m}^\mu={{1}\over{\sqrt{2}r}}(0,0,1,-{{i}\over{sin\theta}}),\eqno{(36)}$$ with scalar products (with respect to ${g_{\mu\nu}}_\Lambda$) $$\matrix{ & l & n & m & \bar{m} \cr l & 0 & 1 & 0 & 0 \cr n & 1 & 0 & 0 & 0 & \cr m & 0 & 0 & 0 & -1 \cr \bar{m} & 0 & 0 & -1 & 0}, \eqno{(37)}$$ reproduces $g^{\mu\nu}_\Lambda$ i.e. $$\tilde{g}^{\mu\nu}=g^{\mu\nu}_\Lambda.\eqno{(38)}$$

\

The complexification given by (1), together with $$\R\ni u\longrightarrow \C\ni u=u^\prime+ibcos\theta^\prime, \ u^\prime\in (-\infty,+\infty), \eqno{(39)}$$ $\varphi^\prime=\varphi$, and the prescriptions (4) and (5), lead to the transformed tetrad $$l^{\prime\mu}=\delta^\mu_{r^\prime}=(0,1,0,0), \ n^{\prime\mu}=\delta^\mu_{u^\prime}-{{1}\over{2}}f_{r\Lambda}\delta^\mu_{r^\prime}=(1,-{{1}\over{2}}f_{r\Lambda},0,0),$$ $$m^{\prime\mu}={{1}\over{\sqrt{2}(r^\prime+ibcos\theta^\prime)}}((\delta^\mu_{u^\prime}-\delta^\mu_{r^\prime})ibsin\theta^\prime+\delta^\mu_{\theta^\prime}+\delta^\mu_{\varphi^\prime}{{i}\over{sin\theta^\prime}})={{1}\over{\sqrt{2}(r^\prime+ibcos\theta^\prime)}}(ibsin\theta^\prime,-ibsin\theta^\prime,1,{{i}\over{sin\theta^\prime}}),$$ $$\bar{m}^{\prime\mu}=\bar{m^{\prime\mu}}, \eqno{(40)}$$ with $$f_{r\Lambda}=1-{{\Lambda r^{\prime 4}}\over{3\Sigma}},\eqno{(41)}$$ $$\Sigma=r^{\prime 2}+b^2cos^2\theta^\prime,\eqno{(42)}$$ and inverse metric $$\tilde{g}^{\prime\mu\nu}=g^{\prime\mu\nu}_\Lambda\equiv g^{\mu\nu}_{r\Lambda}=(l^{\prime\mu}n^{\prime\nu}+l^{\prime\nu}n^{\prime\mu})-(m^{\prime\mu}\bar{m}^{\prime\nu}+m^{\prime\nu}\bar{m}^{\prime\mu}))$$ $$=\pmatrix{-{{b^2sin^2\theta}\over{\Sigma}} & {{r^{\prime 2}+b^2}\over{\Sigma}} & 0 & -{{b}\over{\Sigma}}\cr \cdot & -{{r^{\prime 2}+b^2-\Lambda r^{\prime 4}/3}\over{\Sigma}} & 0 & {{b}\over{\Sigma}}\cr \cdot & \cdot & -{{1}\over{\Sigma}} & 0 \cr \cdot & \cdot & \cdot & -{{1}\over{\Sigma sin^2\theta}}}.\eqno{(43)}$$ Its inverse gives the $rdS$ and the $rAdS$ metrics: $${g_{\mu\nu}}_{r\Lambda}(u^\prime,r^\prime,\theta^\prime,\varphi^\prime)=\pmatrix{1-{{\Lambda r^{\prime 4}}\over{3\Sigma}} & 1 & 0 & {{b\Lambda sin^2\theta^\prime}\over{3\Sigma}}r^{\prime 4}\cr \cdot & 0 & 0 & -bsin^2\theta^\prime\cr \cdot & \cdot & -\Sigma & 0\cr \cdot & \cdot & \cdot & -{{sin^2\theta^\prime}\over{\Sigma}}A}\eqno{(44)}$$ with $$A=(r^{\prime 2}+b^2)^2-b^2sin^2\theta^\prime(r^{\prime 2}+b^2-\Lambda r^{\prime 4}/3).\eqno{(45)}$$ It is interesting to observe that under the interchange ${{\Lambda r^{\prime 3}}\over{3}}\leftrightarrow 2M$, the Kerr and the $r\Lambda$ metrics go into each other [5] i.e. $${g_{\mu\nu}}_{r\Lambda}(u^\prime,r^\prime,\theta^\prime,\varphi^\prime)\buildrel{{{\Lambda r^{\prime 3}}\over{3}}\leftrightarrow 2M}\over\longleftrightarrow {g_{\mu\nu}}_K(u^\prime,r^\prime,\theta^\prime,\varphi^\prime),\eqno{(46)}$$ where ${g_{\mu\nu}}_K$ is given by (49) with $\Lambda=0$.

\

{\bf 5. Kerr-de Sitter ($KdS$) and Kerr-anti de Sitter ($KAdS$) metrics}

\

The same complexification and change of coordinates and tetrads used in section 4., produce the change $$f_{S\Lambda}=1-{{2M}\over{r}}-{{\Lambda r^2}\over{3}}\longrightarrow f_{K\Lambda}=1-{{2Mr^\prime}\over{\Sigma}}-{{\Lambda {r^\prime}^4}\over{3\Sigma}}\eqno{(47)}$$ and the inverse Kerr-de Sitter (anti de Sitter) metrics $$g^{\mu\nu}_{K\Lambda}(u^\prime,r^\prime,\theta^\prime,\varphi^\prime)=\pmatrix{-{{b^2sin^2\theta}\over{\Sigma}} & {{{r^\prime}^2+b^2}\over{\Sigma}} & 0 & -{{b}\over{\Sigma}}\cr \cdot & -{{{r^\prime}^2+b^2-2Mr^\prime-{{\Lambda {r^\prime}^4}\over{3}}}\over{\Sigma}} & 0 & {{b}\over{\Sigma}}\cr \cdot & \cdot & -{{1}\over{\Sigma}} & 0 \cr \cdot & \cdot & \cdot & -{{1}\over{\Sigma sin^2\theta^\prime}}\cr} \eqno{(48)}$$ with inverse $${g_{\mu\nu}}_{K\Lambda}(u^\prime,r^\prime,\theta^\prime,\varphi^\prime)=\pmatrix{1-{{2Mr^\prime+{{\Lambda {r^\prime}^4}\over{3}}}\over{\Sigma}} & 1 & 0 & {{bsin^2\theta^\prime}\over{\Sigma}}(2Mr^\prime+{{\Lambda {r^\prime}^4}\over{3}})\cr \cdot & 0 & 0 & -bsin^2\theta\cr \cdot & \cdot & -\Sigma & 0 \cr \cdot & \cdot & \cdot & -{{sin^2\theta}\over{\Sigma}}A_K\cr} \eqno{(49)}$$ with $$A_K=({r^\prime}^2+b^2)^2-b^2sin^2\theta^\prime({r^\prime}^2+b^2-2Mr^\prime-{{\Lambda{r^\prime}^4}\over{3}}).\eqno{(50)}$$ The inverse metric (48) is nothing but the inverse metric (43) with the addition of the term $-2Mr^\prime$ in the numerator of $-g^{r^\prime r^\prime}$.

\

{\bf 6. Kerr-Newman-de Sitter ($KNdS$) and Kerr-Newman-anti de Sitter ($KNAdS$) metrics}

\

{\it 6.1. Eddington-Finkelstein coordinates} 

\

Incorporating to $f_{S\Lambda}$ the Reissner-Nordstrom ($RN$) term ${{Q^2}\over{{r^\prime}^2}}$, $Q^2=g^2+q^2$ with $q$: electric charge and $g$: magnetic Dirac charge, $[q]=[g]=[L]$ in geometric units, defines $$f_{RN\Lambda}=1-{{2M}\over{r}}+{{Q^2}\over{r^2}}-{{\Lambda r^2}\over{3}}.\eqno{(51)}$$ To simplify, take $g=0$; then the gauge potential 1-form associated to $q$ is $$A={{q}\over{r}}dt={{q}\over{r}}(du+{{dr}\over{{f_{RN\Lambda}}}})=A_u(r)du+A_r(r)dr.\eqno{(52)}$$ The gauge transformation [11] $A^\prime_\mu(r)=A_\mu(r)+\partial_\mu\psi(r)$ allows us to fix $A^\prime_r(r)=0$ with $A^\prime_u(r)=A_u(r)={{q}\over{r}}$ and contravariant components $$(A^u,A^r,A^\theta,A^\varphi)=(0,{{q}\over{r}},0,0)={{q}\over{r}}l^\mu.\eqno{(53)}$$ Then, following the same strategy as in section 5., one obtains the inverse Kerr-Newman-de Sitter (anti de Sitter) metrics $$g^{\mu\nu}_{KN\Lambda}(u^\prime,r^\prime,\theta^\prime,\varphi^\prime)=\pmatrix{-{{b^2sin^2\theta}\over{\Sigma}} & {{{r^\prime}^2+b^2}\over{\Sigma}} & 0 & -{{b}\over{\Sigma}}\cr \cdot & -{{{r^\prime}^2+b^2-2Mr^\prime +Q^2-{{\Lambda {r^\prime}^4}\over{3}}}\over{\Sigma}} & 0 & {{b}\over{\Sigma}}\cr \cdot & \cdot & -{{1}\over{\Sigma}} & 0 \cr \cdot & \cdot & \cdot & -{{1}\over{\Sigma sin^2\theta^\prime}}\cr} \eqno{(54)}$$ with inverse $${g_{\mu\nu}}_{KN\Lambda}(u^\prime,r^\prime,\theta^\prime,\varphi^\prime)=\pmatrix{1-{{2Mr^\prime -Q^2+{{\Lambda {r^\prime}^4}\over{3}}}\over{\Sigma}} & 1 & 0 & {{bsin^2\theta^\prime}\over{\Sigma}}(2Mr^\prime -Q^2+{{\Lambda {r^\prime}^4}\over{3}})\cr \cdot & 0 & 0 & -bsin^2\theta\cr \cdot & \cdot & -\Sigma & 0 \cr \cdot & \cdot & \cdot & -{{sin^2\theta}\over{\Sigma}}A_{KN}\cr} \eqno{(55)}$$ with $$A_{KN}=({r^\prime}^2+b^2)^2-b^2sin^2\theta^\prime({r^\prime}^2+b^2-2Mr^\prime +Q^2-{{\Lambda{r^\prime}^4}\over{3}}),\eqno{(56)}$$ and the gauge vector $A^{\prime\mu}={{qr^\prime}\over{\Sigma}}\delta^\mu_{r^\prime}$ with covariant components $A^\prime_\mu={g_{\mu\nu}}_{KN\Lambda}A^{\prime\nu}={{qr^\prime}\over{\Sigma}}(1,0,0,-bsin^2\theta^\prime)$ i.e. $$A^\prime={{qr^\prime}\over{\Sigma}}(du^\prime-bsin^2\theta^\prime d\varphi^\prime).\eqno{(57)}$$

\

{\it 6.2. Boyer-Lindquist (B-L) coordinates}

\

The {\it B-L} coordinates $(t,r,\theta,\phi)$ [10] are defined by $$du^\prime=dt+xdr, \ d\varphi^\prime=d\phi+ydr, \ r^\prime=r, \ \theta^\prime=\theta\eqno{(58)}$$ with the condition that the coefficients of $dtdr$ and $drd\phi$ vanish. The result is $$x=-{{r^2+b^2}\over{\Delta}}, \ y=-{{b}\over{\Delta}}$$ with $$ \Delta=r^2+b^2-2Mr+Q^2-{{\Lambda r^4}\over{3}},\eqno{(59)}$$ and the metric $${g_{\mu\nu}}_{KN\Lambda}(t,r,\theta,\phi)=\pmatrix{{{\Delta-b^2sin^2\theta}\over{\Sigma}} & 0 & 0 & {{bsin^2\theta}\over{\Sigma}}(r^2+b^2-\Delta)\cr \cdot & -{{\Sigma}\over{\Delta}} & 0 & 0 \cr \cdot & \cdot & -\Sigma & 0 \cr \cdot & \cdot & \cdot & -{{sin^2\theta}\over{\Sigma}}({(r^2+b^2)}^2-b^2sin^2\theta\Delta)\cr},\eqno{(60)}$$ with inverse $${g^{\mu\nu}}_{KN\Lambda}(t,r,\theta,\phi)=\pmatrix{{{r^2+b^2+b^2sin^2\theta{{(2Mr-Q^2+{{\Lambda r^4}\over{3}})}\over{\Sigma}}}\over{\Delta}} & 0 & 0 & {{b(2Mr-Q^2+{{\Lambda r^4}\over{3}}}\over{\Sigma\Delta}}\cr \cdot & -{{\Delta}\over{\Sigma}} & 0 & 0\cr \cdot & \cdot & -{{1}\over{\Sigma}} & 0\cr \cdot & \cdot & \cdot & -{{\Delta-b^2sin^2\theta}\over{\Sigma\Delta sin^2\theta}}\cr}.\eqno{(61)}$$

\

From these expressions, following the lines of the diagram (62) below, one obtains the $B-L$ form of the metrics for the indicated spaces: $$\matrix{rQ(A)dS\cr M=0\nwarrow \cr RN\buildrel{b=0}\over\longleftarrow KN \buildrel{\Lambda=0}\over\longleftarrow KN(A)dS & \buildrel{b=0}\over\longrightarrow RN(A)dS\cr Q^2=0\swarrow\cr K(A)dS & \buildrel{b=0}\over\longrightarrow S(A)dS\buildrel{M=0}\over\longrightarrow (A)dS\buildrel{\Lambda=0}\over\longrightarrow Mink\cr M=0\downarrow\cr r(A)dS\cr},\eqno{(62)}$$ where $rQ(A)dS$ is a rotating charged anti de Sitter or de Sitter cosmological universe, $KN$ is the Kerr-Newman metric, and $Mink$ denotes Minkowski space.

\

In $B-L$ coordinates, the gauge potential is given by $$A^\prime={{qr}\over{\Sigma}}(dt-{{\Sigma}\over{\Delta}}dr-bsin^2\theta d\phi).\eqno{(63)}$$ Again, the term proportional to $dr$ can be set equal to zero through a gauge transformation since $A^\prime_r=-{{qr}\over{\Delta(r)}}=A^\prime_r(r)$, and one ends with the usual form $$A^\prime={{qr}\over{\Sigma}}(dt-bsin^2\theta d\phi). \eqno{(64)}$$ For the electromagnetic (electric) field tensor one has: 

\

{\it Covariant components}: $$F_{\mu\nu}=D_\mu A_\nu-D_\nu A_\mu=\partial_\mu A_\nu-\partial_\nu A_\mu=\pmatrix{0 & F_{tr} & F_{t\theta} & 0 \cr -F_{tr} & 0 & 0 & F_{r\phi}\cr -F_{t\theta} & 0 & 0 & F_{\theta\phi}\cr 0 & -F_{r\phi} & -F_{\theta\phi} & 0\cr},\eqno{(65)}$$ with $$F_{tr}=q{{2r^2-\Sigma}\over{\Sigma^2}}, \ F_{t\theta}=-{{qb^2rsin(2\theta)}\over{\Sigma^2}}, \ F_{r\phi}=bsin^2\theta F_{tr}, \ F_{\theta\phi}=-{{qbrsin(2\theta)(r^2+b^2)}\over{\Sigma^2}},$$ $$[F_{tr}]=[L]^{-1}, \ [F_{t\theta}]=[L]^0, \ [F_{\theta\phi}]=[L]^1, \ [F_{r\phi}]=[L]^0;\eqno{(66)}$$

\

{\it Contravariant components}: $$F^{\mu\nu}={g^{\mu\rho}}_{KN\Lambda}{g^{\nu\sigma}}_{KN\Lambda}F_{\rho\sigma}=\pmatrix{0 & F^{tr} & F^{t\theta} & 0\cr -F^{tr} & 0 & 0 & F^{r\phi} \cr -F^{t\theta} & 0 & 0 & F^{\theta\phi}\cr 0 & -F^{r\phi} & -F^{\theta\phi} & 0 \cr},\eqno{(67)}$$ with $$F^{tr}=-{{q(2r^2-\Sigma)(r^2+b^2)}\over{\Sigma^3}}, \ F^{t\theta}={{qb^2rsin(2\theta)}\over{\Sigma^3}}, \ F^{r\phi}={{qb(2r^2-\Sigma)}\over{\Sigma^3}}, \ F^{\theta\phi}=-{{qbrsin(2\theta)}\over{\Sigma^3sin^2\theta}},$$ $$ [F^{tr}]=[L]^{-1}, \ [F^{t\theta}]=[L]^{-2}, \ [F^{r\phi}]=[L]^{-2}, \ [F^{\theta\phi}]=[L]^{-3}; \eqno{(68)}$$ 

\

{\it Mixed components}: $${F^\mu}_\nu={g^{\mu\rho}}_{KN\Lambda}F_{\rho\nu}=\pmatrix{0 & {F^t}_r & {F^t}_\theta & 0\cr {F^r}_t & 0 & 0 & {F^r}_\phi\cr {F^\theta}_t & 0 & 0 & {F^\theta}_\phi\cr 0 & {F^\phi}_r & {F^\phi}_\theta & 0\cr},\eqno{(69)}$$ with $${F^t}_r=g^{tt}F_{tr}+g^{t\phi}F_{\phi r}={{q(2r^2-\Sigma)(r^2+b^2)}\over{\Sigma^2\Delta}}, \ [{F^t}_r]=[L]^{-1},\eqno{(70)}$$ $${F^t}_\theta=g^{tt}F_{t\theta}+g^{t\phi}F_{\phi\theta}=-{{qb^2rsin(2\theta)}\over{\Sigma^2}}, \ [{F^t}_\theta]=[L]^0, \eqno{(71)}$$ $${F^r}_t=g^{rr}F_{rt}={{q\Delta(2r^2-\Sigma)}\over{\Sigma^3}}, \ [{F^r}_t]=[L]^{-1}, \eqno{(72)}$$ $${F^\theta}_t=g^{\theta\theta}F_{\theta t}=-{{qb^2rsin(2\theta)}\over{\Sigma^3}}, \ [{F^\theta}_t]=[L]^{-2},\eqno{(73)}$$ $${F^r}_\phi=g^{rr}F_{r\phi}=-{{qb\Delta sin^2\theta(2r^2-\Sigma)}\over{\Sigma^3}}, \ [{F^r}_\phi]=[L]^0, \eqno{(74)}$$ $${F^\phi}_r=g^{\phi t}F_{tr}+g^{\phi\phi}F_{\phi r}={{qb(2r^2-\Sigma)}\over{\Sigma^2\Delta}}, \ [{F^\phi}_r]=[L]^{-2},\eqno{(75)}$$ $${F^\theta}_\phi=g^{\theta\theta}F_{\theta\phi}={{qbrsin(2\theta)(r^2+b^2)}\over{\Sigma^3}}, \ [{F^\theta}_\phi]=[L]^{-1},\eqno{(76)}$$ $${F^\phi}_\theta=g^{\phi t}F_{t\theta}+g^{\phi\phi}F_{\phi\theta}=-{{qbrsin(2\theta)}\over{\Sigma^2sin^2\theta}}, \ [{F^\phi}_\theta]=[L]^{-1}.\eqno{(77)}$$

\

These expressions allow us to compute the electromagnetic part of the energy-momentum tensor $T_{\mu\nu}^E$: $$T_{\mu\nu}^E={{1}\over{4\pi}}(F_{\mu\rho}{F^\rho}_\nu+{{1}\over{4}}g_{\mu\nu}F_{\rho\sigma}F^{\rho\sigma})=\pmatrix{T_{tt} & 0 & 0 & T_{t\phi}\cr \cdot & T_{rr} & 0 & 0 \cr \cdot & \cdot & T_{\theta\theta} & 0 \cr \cdot & \cdot & \cdot & T_{\phi\phi}\cr}$$ $$={{q^2}\over{8\pi\Sigma}}\pmatrix{{{\Delta+b^2sin^2\theta}\over{\Sigma^2}} & 0 & 0 & {{-bsin^2\theta}\over{\Sigma^2}}A_{t\phi}\cr \cdot & -{{1}\over{\Delta}} & 0 & 0 \cr \cdot & \cdot & 1 & 0 \cr \cdot & \cdot & \cdot & A_{\phi\phi}\cr},\eqno{(78)}$$ with $$A_{t\phi}=r^2+b^2+\Delta\eqno{(79)}$$ and $$A_{\phi\phi}={{sin^2\theta}\over{\Sigma^2}}((r^2+b^2)^2+b^2\Delta sin^2\theta).\eqno{(80)}$$ Clearly, for $q=0$, $T_{\mu\nu}^E=0$ independently of the values for $M$, $b$ and $\Lambda$, while for $q\neq 0$, $b=0$ and $\Lambda=0$ one recovers the $RN$ energy-momentum tensor.

\

{\bf 7. Conclusion}

\

A simple trick (eq.(2)) which unifies the usual treatment of the $r^{-1}$ and $r^{-2}$ terms after complexification of the radial coordinate $r$ in the Newman-Janis approach to the Kerr and Kerr-Newman metrics, allows us to consider under the same scheme terms proportional to $r^2$ appearing in the de Sitter ($dS$) and anti de Sitter ($AdS$) cases ({\it cf.} ref. [11], eq.(2.6c)). In particular, for the massive rotating cosmological cases ($K(A)dS$ and $KN(A)dS$) our solution, eq.(55), (or (49) for $Q^2=0$), coincides with that of Ibohal [6], eq.(6.41), but, as is the case of this author, is different from those of Carter [12], Gibbons and Hawking [13], Mallett [14], Koberlein [15], and others. Some details of the calculations and a complete study of the geometry associated with the metric (60), like Kruskal coordinates, Penrose diagram, horizons, ergospheres, etc., and the complete energy-momentum tensor, will be published elsewhere.

\

{\bf Acknowledgment}

\

This work was partially supported by the project PAPIIT IN105413, DGAPA-UNAM, M\'exico.

\

{\bf References}

\

[1] Newman, E.T. and Janis, A.I. (1965). Journal of Mathematical Physics {\bf 6}, 915-917.

\

[2] Drake, S.P. and Szekeres, P. (2000). General relativity and Gravitation {\bf 32}, 445-457.

\

[3] Akcay, S. and Matzner, R. (2011). Classical and Quantum Gravity {\bf 28}, 085012.

\

[4] Peng, Z. (2009). DAMPT, Center for Mathematical Physics, Cambridge.

\

[5] Azreg-Ainou, M. (2014). Physics Letters B {\bf 730}, 95-98.

\

[6] Ibohal, N. (2005). General Relativity and Gravitation {\bf 37}, 19-51.

\

[7] Socolovsky, M. (2014). Annales de la Fondation Louis de Broglie {\bf 39}, 1-49.

\

[8] Deser, S. and Levin, O. (1999). Physical Review D {\bf 59}, 064004.

\

[9] Reine, D. and Thomas, E. (2010). Black Holes. An Introduction, 2nd. edition, Imperial College Press, London.

\

[10] Boyer, R.H. and Lindquist, R.W. (1967). Journal of Mathematical Physics {\bf 8}, 265-281.

\

[11] Erbin, H. (2014). arXiv: gr-qc/1410.2602v1.

\

[12] Carter, B. (1973). Les Astre Occlus, Gordon and Breach, New York, 57-124.

\

[13] Gibbons, G.W. and Hawking, S.W. (1977). Physical Review D {\bf 15}, 2738-2751.

\

[14] Mallett, R.L. (1988). Physics Letters A {\bf 126}, 226-228.

\

[15] Koberlain, B.D. (1995). Physical Review D {\bf 51}, 6783-6787.

\

\

\

*{\it Postdoctoral fellowship, CONICET, Argentina}

\

**{\it With a leave of absence from Instituto de Ciencias Nucleares, Universidad Nacional Aut\'onoma de M\'exico, M\'exico}

\

e-mails: egurreta@ungs.edu.ar, msocolov@ungs.edu.ar, socolovs@nucleares.unam.mx

\end

\centerline{\bf NEWMAN-JANIS ALGORITHM REVISITED}

\

\

\centerline{O. Brauer and H. A. Camargo}
\centerline{\it Facultad de Ciencias, Universidad Nacional Aut\'onoma de M\'exico}
\centerline{\it Circuito Exterior, Ciudad Universitaria, 04510, M\'exico D. F., M\'exico} 
\centerline{and}
\centerline{M. Socolovsky}
\centerline{\it  Instituto de Ciencias Nucleares, Universidad Nacional Aut\'onoma de M\'exico}
\centerline{\it Circuito Exterior, Ciudad Universitaria, 04510, M\'exico D. F., M\'exico} 

\

{\it The purpose of the present article is to show that the Newman-Janis and Newman et al algorithm used to derive the Kerr and Kerr-Newman metrics respectively, automatically leads to the extension of the initial non negative polar radial coordinate $r$ to a cartesian coordinate $r^\prime$ running from $-\infty$ to $+\infty$, thus introducing in a natural way the region $-\infty<r^\prime<0$ in the above spacetimes. Using Boyer-Lindquist and ellipsoidal coordinates, we discuss some geometrical aspects of the positive and negative regions of $r^\prime$, like horizons, ergosurfaces, and foliation structures.}

\

{\bf 1. Introduction}

\

The uncharged Kerr ($K$) (Kerr, 1963) and the charged Kerr-Newman ($KN$) (Newman et al, 1965) axially symmetic stationary spacetimes have, in contradistinction to the charged static spherical Reisner-Nordstrom ($RN$) solution (Reissner, 1916; Nordstrom, 1918), an asymptotically flat region which involves, both in the Eddington-Finkelstein ($EF$) (Eddington, 1924; Finkelstein, 1958) and  Boyer-Lindquist ($BL$) (Boyer and Lindquist, 1967) coordinates, a radial cordinate taking not only positive and zero values, but also negative ones, form 0 to -$\infty$. This strange situation is usually explained by the fact that, since the curvature singularity of both the $K$ and $KN$ solutions is in the circular boundary of a non singular open disk in the equatorial plane, the spacetimes can be continued through it, to regions in which the radial coordinate becomes negative (regions $III$ and $III^\prime$ in the Penrose-Carter diagram (Penrose, 1963; Carter, 1966), section {\bf 8}). As it stands, the argument, though correct, does not imply the necessity of this continuation, but only it allows for its possibility. 

\

In this note we show that the complexification procedure involved in the Newman-Janis algorithm ($NJA$) (Newman and Janis, 1965) used to derive the $K$ and $KN$ metrics, automatically leads to the extension to negative values of the originally non negative polar radial coordinate, leaving from the outset with coordinates $(u^\prime,r^\prime,\theta,\varphi)$ (or $(t^\prime,r^\prime, \theta,\phi)$) taking values in $\R^2\times S^2$ i.e. two cartesian ($u^\prime,r^\prime$ or $t^\prime,r^\prime$) and two compact ($\theta,\varphi$ or $\theta,\phi$) coordinates ($u^\prime$ is the retarded $EF$ time and $t$ the $BL$ time). Though at the epoque of its inception and for many years, the Newman-Janis and Newman et al derivations respectively of the $K$ and $KN$ metrics were considered as flukes, recent work by Drake and Szekeres (Drake and Szekeres, 2000) has put the algorithm on a more solid ground by proving uniqueness theorems for the kind of solutions which can be derived using the algorithm. 

\

In sections {\bf 2} to {\bf 5} we review the $NJA$ derivation of the $K$ and $KN$ metrics emphasizing, in section {\bf 4}, how the complexification of the radial coordinate automatically implies the range $(-\infty,+\infty)$ for its real part $r^\prime$. Using $BL$ (section {\bf 6}) and ellipsoidal coordinates, in section {\bf 7} we exhibit the horizons, ergosurfaces, and foliation structures of the $KN$ and $K$ solutions (respectively for the cases $M^2>a^2+Q^2$ and $M^2>a^2$) in both regions $r^\prime >0$ and $r^\prime <0$, and give a brief discussion of the spatial topology of these constructions. For completeness, the basic cell of the Penrose-Carter diagram of the $K$ and $KN$ spacetimes is exhibited in section {\bf 8}.  

\

{\bf 2. Reissner-Nordstrom spacetime} 

\

Our starting point is the Reissner-Nordstrom ($RN$) spacetime written in terms of the retarded Eddington
-Finkelstein coordinates $(u,r,\theta,\varphi)$ with $u\in\R=(-\infty,+\infty)$, $r\in\R_{\geq}=[0,+\infty)$, and $\theta,\varphi\in S^2$ i.e. $\theta\in[0,\pi]$ and $\varphi\in [0,2\pi)$: $$ds^2_{RN}=fdu^2+2dudr-r^2d^2\Omega, \ \ d^2\Omega=d\theta^2+sin^2\theta d\varphi^2 \eqno{(1)}$$ with $$f={{r^2-2Mr+Q^2}\over{r^2}}=1-{{2M}\over{r}}+{{Q^2}\over{r^2}}, \ \ [f]=[L]^0, \eqno{(2)}$$ where $M$ is the gravitating (positive) mass and $Q^2=q^2+p^2$ where $q$ is the electric charge and $p$ is the hypotetical abelian (Dirac) magnetic charge. The metric corresponding to (1) is $${g_{\mu\nu}}_{RN}=\pmatrix{f&1&0&0\cr 1&0&0&0\cr 0&0&-r^2&0\cr 0&0&0& -r^2sin^2\theta\cr}, \ \ det {g_{\mu\nu}}_{RN}=-r^4sin^2\theta,\eqno{(3)}$$ with inverse $${g^{\mu\nu}}_{RN}=\pmatrix{0&1&0&0\cr 1&-f&0&0\cr 0&0&-r^{-2}&0\cr 0&0&0&-r^{-2}sin^{-2}\theta\cr}.\eqno{(4)}$$ ($\mu=0,1,2,3$ respectively correspond to $u,r,\theta,\varphi.$)

\

{\bf 3. Null tetrad}

\

At each point of the $RN$ manifold we can choose as a basis of the corresponding tangent space the null tetrad consisting of the following linear independent 4-vectors:

\

$$l=l^\mu{{\partial}\over{\partial x^\mu}}=\delta^\mu_1{{\partial}\over{\partial x^\mu}}={{\partial}\over{\partial x^1}}={{\partial}\over{\partial r}},\eqno{(5a)}$$ $$n=n^\mu{{\partial}\over{\partial x^\mu}}=(\delta^\mu_0-{{f}\over{2}}\delta^\mu_1){{\partial}\over{\partial x^\mu}}={{\partial}\over{\partial u}}-{{f}\over{2}}{{\partial}\over{\partial r}},\eqno{(5b)}$$ $$m=m^\mu{{\partial}\over{\partial x^\mu}}={{1}\over{\sqrt{2}r}}(\delta^\mu_2+{{i}\over{sin\theta}}\delta^\mu_3){{\partial}\over{\partial x^\mu}}={{1}\over{\sqrt{2}r}}({{\partial}\over{\partial\theta}}+{{i}\over{sin\theta}}{{\partial}\over{\partial\varphi}}),\eqno{(5c)}$$ $$\bar{m}=\bar{m}^\mu{{\partial}\over{\partial x^\mu}}={{1}\over{\sqrt{2}r}}(\delta^\mu_2-{{i}\over{sin\theta}}\delta^\mu_3){{\partial}\over{\partial x^\mu}}={{1}\over{\sqrt{2}r}}({{\partial}\over{\partial\theta}}-{{i}\over{sin\theta}}{{\partial}\over{\partial\varphi}}),\eqno{(5d)}$$ i.e. $$l^\mu=(l^u,l^r,l^\theta,l^\varphi)=(0,1,0,0),\eqno{(5a')}$$ $$n^\mu=(n^u,n^r,n^\theta,n^\varphi)=(1,-{{f}\over{2}},0,0),\eqno{(5b')}$$ $$m^\mu=(m^u,m^r,m^\theta,m^\varphi)={{1}\over{\sqrt{2}r}}(0,0,1,{{i}\over{sin\theta}}),\eqno{(5c')}$$ $$\bar{m}^\mu=(\bar{m}^u,\bar{m}^r,\bar{m}^\theta,\bar{m}^\varphi)={{1}\over{\sqrt{2}r}}(0,0,1,-{{i}\over{sin\theta}}),\eqno{(5d')}$$ with covariant components $b_\mu={g_{\mu\nu}}_{RN}b^\nu$ given by $$l_\mu=(1,0,0,0,),\eqno{(5a'')}$$ $$n_\mu=({{f}\over{2}},1,0,0),\eqno{(5b'')}$$ $$m_\mu={{1}\over{\sqrt{2}r}}(0,0,-r^2,-ir^2sin\theta),\eqno{(5c'')}$$ $$\bar{m}_\mu={{1}\over{\sqrt{2}r}}(0,0,-r^2,ir^2sin\theta).\eqno{(5d'')}$$ The scalar products $a\cdot b={g_{\mu\nu}}_{RN}a^\mu b^\nu$ of the members of the tetrad are given in the following table: $$\matrix{a\cdot b&l&n&m&\bar{m}\cr l&0&1&0&0\cr n&1&0&0&0\cr m&0&0&0&-1\cr \bar{m}&0&0&-1&0\cr}$$

\

\centerline{Table I}

\

{\it Remark}: It is not strange that for the tangent spaces of a real manifold (in this case 4-dimensional) one chooses a basis containing complex tangent vectors; consider e.g. the basis of the Lie algebra of the $SU(2)$ group: $\tau_k={{i\sigma_k}\over{2}}, \ k=1,2,3.$ The real character of the tangent spaces depends on the field of numbers, not on the basis vectors.

\

It is an easy exercise to verify that the quantity $$\tilde{g}^{\mu\nu}=(l^\mu n^\nu+l^\nu n^\mu)-(m^\mu\bar{m}^\nu+m^\nu\bar{m}^\mu)\eqno{(6)}$$ is nothing but the inverse $RN$ metric: $$\tilde{g}^{\mu\nu}={g^{\mu\nu}}_{RN}.\eqno{(7)}$$ It is interesting to note that if the tetrad is denoted by $$(e_1,e_2,e_3,e_4)=(l,n,m,\bar{m}),\eqno{(8)}$$ then $$\tilde{g}^{\mu\nu}={e_a}^\mu J^{ab}{e_b}^\nu\eqno{(9)}$$ with $$J=\pmatrix{\sigma_1&0\cr 0&-\sigma_1\cr},\eqno{(10)}$$ where $\sigma_1=\pmatrix{0&1\cr 1&0\cr}$ is the Pauli matrix. $J$ is related to the $\alpha^k=\pmatrix{0&\sigma_k\cr \sigma_k&0\cr}$, $k=1,2,3$, matrices of the standard representation of the Dirac equation, with $\sigma_2=\pmatrix{0&-i\cr i&0}$ and $\sigma_3=\pmatrix{1&0\cr 0&-1\cr}$, through the similarity transformations $$S_k^{-1}JS_k=\alpha^k, \ S_1=\pmatrix{1&1\cr 1&-1\cr}, \ S_2=\pmatrix{1&\sigma_1\sigma_2\cr 1&-\sigma_1\sigma_2\cr}, \ S_3=\pmatrix{1&\sigma_1\sigma_3\cr 1&-\sigma_1\sigma_3},\eqno{(11)}$$ with $$S_1^{-1}={{1}\over{2}}\pmatrix{1&1\cr 1&-1}, \ S_2^{-1}={{1}\over{2}}\pmatrix{1&1\cr -i\sigma_3&i\sigma_3\cr}, \ S_3^{-1}={{1}\over{2}}\pmatrix{1&1\cr i\sigma_2&-i\sigma_2}.\eqno{(12)}$$  

\

{\bf 4. Complexification}

\

The following step is the crucial element of the NJA: {\it the coordinates} $r$ {\it and} $u$ {\it are complexified} and a new real positive parameter $a$ (later identified with the angular momentum per unit mass) is introduced:

\

$$r\in\R_{\geq}\longrightarrow r\in \C, \ \ r=r^\prime-iacos\theta,\eqno{(13)}$$ $$u\in\R \longrightarrow u\in\C, \ \ u=u^\prime+iacos\theta, \eqno{(14)}$$ with $\bar{r}=r^\prime+iacos\theta$ and $\bar{u}=u^\prime-iacos\theta$. But now $$r^\prime\in(-\infty,+\infty), \eqno{(15)}$$ i.e. $r^\prime$ has become a {\it cartesian coordinate}. In particular, this will imply that when dealing with the Boyer-Lindquist coordinates in section {\bf 6}, no appeal for an analytic continuation to the asymptotically flat (A.F.) region $r^\prime<0$ through the open disk $y^2+z^2<a^2$ where no singularity occurs will be required: implicitly, this analytic continuation was already done through the complexification (13). The domains of definition of $r$ and $u$ are shown in Figure 1. Clearly, $[a]=[L]^1$. 

\

The new set of real coordinates is $(u^\prime,r^\prime,\theta^\prime,\varphi^\prime)$ with $u^\prime\in(-\infty,+\infty)$, $\theta^\prime=\theta$, and $\varphi^\prime=\varphi$. The total domain of these coordinates is $\R^2\times S^2$.

\

Since in the end one needs a real spacetime, $f$ must remain real and so its change is given by $$f(r)\longrightarrow f(r,\bar{r})=1-M({{1}\over{r}}+{{1}\over{\bar{r}}})+{{Q^2}\over{r\bar{r}}}=1-{{2Mr^\prime-Q^2}\over{\Sigma}}\eqno{(16)}$$ where $$\Sigma={r^\prime}^2+a^2cos^2\theta.\eqno{(17)}$$ ($[\Sigma]=[L]^2$.) The transformation of the tetrad is $${e_a}^\mu\longrightarrow {e_a^\prime}^{\mu}={{\partial{x^\prime}^\mu}\over{\partial x^\nu}}{e_a}^\nu\equiv ({l^\prime}^\mu,{n^\prime}^\mu,{m^\prime}^\mu,{\bar{m^\prime}}^\mu)\eqno{(18)}$$ with $$\pmatrix{{{\partial u^\prime}\over{\partial u}} & {{\partial u^\prime}\over{\partial r}} & {{\partial u^\prime}\over{\partial \theta}} & {{\partial u^\prime}\over{\partial \varphi}} \cr {{\partial r^\prime}\over{\partial u}} & {{\partial r^\prime}\over{\partial r}} & {{\partial r^\prime}\over{\partial \theta}} & {{\partial r^\prime}\over{\partial \varphi}} \cr {{\partial \theta^\prime}\over{\partial u}} & {{\partial \theta^\prime}\over{\partial r}} & {{\partial \theta^\prime}\over{\partial \theta}} & {{\partial \theta^\prime}\over{\partial \varphi}} \cr {{\partial \varphi^\prime}\over{\partial u}} & {{\partial \varphi^\prime}\over{\partial r}} & {{\partial \varphi^\prime}\over{\partial \theta}} & {{\partial \varphi^\prime}\over{\partial \varphi}}}=\pmatrix{1&0&iasin\theta&0\cr 0&1&-iasin\theta&0\cr 0&0&1&0\cr 0&0&0&1\cr}\eqno{(19)}$$ and $$l^\nu=\delta^\nu_1, \ n^\nu=\delta^\nu_0-{{1}\over{2}}f(r,\bar{r})\delta^\nu_1, \ m^\nu={{1}\over{\sqrt{2}\bar{r}}}(\delta^\nu_2+{{i}\over{sin\theta}}\delta^\nu_3), \ \bar{m^\nu}={{1}\over{\sqrt{2}r}}(\delta^\nu_2-{{i}\over{sin\theta}}\delta^\nu_3).\eqno{(20)}$$ One obtains: $${l^\prime}^\mu=\delta^\mu_1, \eqno{(21a)}$$ $${n^\prime}^\mu=\delta^\mu_0-{{1}\over{2}}(1-{{2Mr^\prime-Q^2}\over{\Sigma}})\delta^\mu_1, \eqno{(21b)}$$ $${m^\prime}^\mu={{1}\over{\sqrt{2}(r^\prime+iacos\theta)}}((\delta^\mu_0-\delta^\mu_1)iasin\theta+\delta^\mu_2+\delta^\mu_3{{i}\over{sin\theta}}),\eqno{(21c)}$$ $${\bar{m^\prime}}^\mu={{1}\over{\sqrt{2}(r^\prime-iacos\theta)}}(-(\delta^\mu_0-\delta^\mu_1)iasin\theta+\delta^\mu_2-\delta^\mu_3{{i}\over{sin\theta}}).\eqno{(21d)}$$

\

{\bf 5. Kerr-Newman metric}

\

The ``miracle", though in some way justified by Drake and Szekeres through the proof of important uniqueness theorems from the NJA for vacuum and Einstein-Maxwell solutions, is that the quantity $${g^\prime}^{\mu\nu}=({l^\prime}^\mu{n^\prime}^\nu+{l^\prime}^\nu{n^\prime}^\mu)-({m^\prime}^\mu\bar{{m^\prime}}^\nu+{m^\prime}^\nu\bar{{m^\prime}}^\mu)\eqno{(22)}$$ is the inverse Kerr-Newman ($KN$) metric in Eddington-Finkelstein retarded coordinates $(u^\prime,r^\prime,\theta,\varphi)$: $${g^\prime}^{\mu\nu}={g^{\mu\nu}}_{KN}. \eqno{(23)}$$ In fact, a straightforward calculation leads to $${g^\prime}^{\mu\nu}=\pmatrix{g^{u^\prime u^\prime}&g^{u^\prime r^\prime}&g^{u^\prime \theta}&g^{u^\prime \varphi}\cr \cdot&g^{r^\prime r^\prime}&g^{r^\prime \theta}&g^{r^\prime \varphi}\cr \cdot&\cdot&g^{\theta \theta}&g^{\theta \varphi}\cr \cdot&\cdot&\cdot&g^{\varphi\varphi}\cr}=\pmatrix{{{-a^2sin^2\theta}\over{\Sigma}}&{{{r^\prime}^2+a^2}\over{\Sigma}}&0&-{{a}\over{\Sigma}}\cr \cdot&-{{\Sigma-2Mr^\prime+Q^2+a^2sin^2\theta}\over{\Sigma}}&0&{{a}\over{\Sigma}}\cr \cdot&\cdot&-{{1}\over{\Sigma}}&0\cr \cdot&\cdot&\cdot&-{{1}\over{\Sigma sin^2\theta}}}\eqno{(24)}$$ with inverse $${g_{\mu\nu}}_{KN}=\pmatrix{1-{{2Mr^\prime-Q^2}\over{\Sigma}}&1&0&asin^2\theta{{2Mr^\prime-Q^2}\over{\Sigma}}\cr \cdot&0&0&-asin^2\theta\cr \cdot&\cdot&-\Sigma&0\cr \cdot&\cdot&\cdot&-sin^2\theta{{A}\over{\Sigma}}\cr}\eqno{(25)}$$ with $$A=\Sigma({r^\prime}^2+a^2)+a^2sin^2\theta{{2Mr^\prime-Q^2}\over{\Sigma}}=({r^\prime}^2+a^2)^2-a^2sin^2\theta\Delta,\eqno{(26)}$$ and $$\Delta={r^\prime}^2+a^2-2Mr^\prime+Q^2.\eqno{(27)}$$ The square of the spacetime element is $$ds^2_{KN}=(1-{{2Mr^\prime-Q^2}\over{\Sigma}})d{u^\prime}^2+2du^\prime dr^\prime+2asin^2\theta{{2Mr^\prime-Q^2}\over{\Sigma}}du^\prime d\varphi-2asin^2\theta dr^\prime d\varphi-\Sigma d\theta^2-sin^2\theta{{A}\over{\Sigma}}d\varphi^2.\eqno{(28)}$$ $ds^2_{KN}$ reduces to $ds^2_{RN}$ for $a=0$ and $r^\prime\geq 0$. 

\

{\bf 6. Boyer-Lindquist coordinates}

\

The change of coordinates $$dt=du^\prime-{{{r^\prime}^2+a^2}\over{\Delta}}dr^\prime, \ \ d\phi=d\varphi-{{a}\over{\Delta}}dr^\prime \eqno{(29)}$$ leads to the Boyer-Lindquist form of the Kerr-Newman spacetime: $$ds^2_{KN}\vert_{BL}={{\Delta-a^2sin^2\theta}\over{\Sigma}}dt^2-{{\Sigma}\over{\Delta}}d{r^\prime}^2-\Sigma d\theta^2-sin^2\theta{{A}\over{\Sigma}}d\phi^2+{{2asin^2\theta}\over{\Sigma}}({r^\prime}^2+a^2-\Delta)dtd\phi\eqno{(30)}$$ i.e. $${g_{\mu\nu}}_{KN}\vert_{BL}=\pmatrix{1-{{2Mr^\prime-Q^2}\over{\Sigma}}&0&0&{{asin^2\theta}\over{\Sigma}}({r^\prime}^2+a^2-\Delta)\cr \cdot&-{{\Sigma}\over{\Delta}}&0&0\cr \cdot&\cdot&-\Sigma&0\cr \cdot&\cdot&\cdot&-sin^2\theta{{A}\over{\Sigma}}\cr}.\eqno{(31)}$$ Horizons $H_+$ and $H_-$ are defined by the zeros of $\Delta$; it is clear that only for $r^\prime >0$ and $M^2\geq a^2+Q^2$ horizons exist, with $$r^\prime_\pm=M\pm\sqrt{M^2-(a^2+Q^2)}.\eqno{(32)}$$ It can be easily seen that for $M^2>a^2+Q^2$, $r^\prime_- <\sqrt{a^2+Q^2}$, in particular $r^\prime_-<a$ for the Kerr case $Q^2=0$; for the extreme cases $M^2=a^2+Q^2$, $r^\prime_-=r^\prime_+=M=\sqrt{a^2+Q^2}$. 

\

For $r^\prime < 0$, $$\Delta={r^\prime}^2+a^2+2M\vert r^\prime\vert+Q^2>0,\eqno{(33)}$$ which has no real roots. The same occurs for the ergosurfaces $S_+$ and $S_-$ whose equations are given by the zeros of ${g_{tt}}_{KN}\vert_{BL}$; for $r^\prime < 0$, $${g_{tt}}_{KN}\vert_{BL}=1+{{2M\vert r^\prime\vert +Q^2}\over{\Sigma}}>1.\eqno{(34)}$$ Also, as is well known, from (17), (26), (27) and (30), $$ds^2_{KN}\vert_{BL}\longrightarrow dt^2-d{r^\prime}^2-{r^\prime}^2(d\theta^2+sin^2\theta d\phi^2)\eqno{(35)}$$ as $r^\prime\to \pm\infty$ i.e. the metric is A.F. in both the $r^\prime>0$ and $r^\prime<0$ regions.

\

{\bf 7. Generalized ellipsoidal coordinates} 

\

As is well known, the use of Kerr-Schild ($KS$) coordinates for the $KN$ metric and their restriction to ellipsoidal coordinates ($EC$) for the Kerr ($K$) metric ($ds^2_K=ds^2_{KN}\vert_{Q^2=0}$), allows to show that both metrics are flat (Minkowskian) for $M^2=Q^2=0$ in the $KN$ case and $M^2=0$ in the $K$ case. The general form of the spatial part of these coordinates allowing for both positive and negative values of $r^\prime$ is $$x_\pm=\sqrt{{r^\prime}^2+a^2} \ sin\theta \ cos(\phi+F(r^\prime)),\eqno {(36a)}$$ $$y_\pm=\sqrt{{r^\prime}^2+a^2} \ sin\theta \ sin(\phi+F(r^\prime)),\eqno {(36b)}$$ $$z_\pm=r^\prime cos\theta \eqno{(36c)}$$ with the + (-) sign corresponding to $r^\prime>0$ ($r^\prime<0$) and $$F(r^\prime)=\lbrace\matrix{-arc \ tan({{a}\over{r^\prime}}), \ KS \cr 0, \ EC}. \eqno{(37)}$$ $x_\pm$, $y_\pm$, and $z_\pm$ are cartesian coordinates defining two $\R^3$ spaces with opposite orientations: right handed for $r^\prime>0$ and left-handed for $r^\prime<0$. In terms of these coordinates, $$ds^2_{KN}\vert_{BL}\vert_{M=0, \ Q^2=0}=ds^2_K\vert_{BL}\vert_{M=0}=dt^2-(dx_\pm^2+dy_\pm^2+dz_\pm^2). \eqno{(38)}$$ In particular, from (30), $$ds^2_K\vert_{BL}\vert_{M=0}=dt^2-({{\Sigma}\over{\Delta}}{dr^\prime}^2+\Sigma d\theta^2+\Delta sin^2\theta d\phi^2).\eqno{(39)}$$ In terms of $(x^\prime,y^\prime,z^\prime)$, $r^\prime$ is given by $$r^\prime=\pm{{1}\over{\sqrt{2}}}\sqrt{(x_\pm^2+y_\pm^2+z_\pm^2-a^2)+\sqrt{(x_\pm^2+y_\pm^2+z_\pm^2-a^2)+4a^2z_\pm^2}}.\eqno{(40)}$$

\

Both the $KS$ and the $EC$ coordinate systems admit, at each $t$, the same foliations of the two $\R^3$ spaces:

\

i) Confocal ellipsoids of revolution $r^\prime=const.$, foci at $x_+^2+y_+^2=a^2$, $z_+=0$: $${{x_\pm^2+y_\pm^2}\over{{r^\prime}^2 +a^2}}+{{z_\pm^2}\over{{r^\prime}^2}}=1,\eqno{(41)}$$ with larger semi-axis =$\sqrt{{r^\prime}^2+a^2}$ and smaller semi-axis=$\vert r^\prime\vert$. For $r^\prime>0$, the ellipsoids corresponding to $r^\prime_\pm$ are the horizons $H_\pm$, embedded in the $(x_+,y_+,z_+)$ space.

\

ii) Confocal 1-sheet hyperboloids of revolution $\theta=const.$, foci at $x_+^2+y_+^2=a^2$, $z_+=0$: $${{x_\pm^2+y_\pm^2}\over{a^2sin^2\theta}}-{{z_\pm^2}\over{a^2cos^2\theta}}=1.\eqno{(42)}$$ The surfaces $\phi=const.$ for the $KS$ coordinates are discussed e.g. in Krasi\'nski and Pleba\'nski (2006), but for the $EC$ coordinates they are simply given by: 

\

iii) Planes through the $z_\pm$ axis.

\

Since the $EC$ system can accommodate the horizons $H_\pm$ and the ergosurfaces $S_\pm$ (see below), we shall restrict the discussion to this coordinate system. 

\

The {\it curvature singularity} of (30) is given by the condition $$\Sigma=0 \eqno {(43a)}$$ which, by (17), implies $$r^\prime=0, \ \theta={{\pi}\over{2}}.\eqno{(43b)}$$ By (36), $r^\prime=0$ defines the disks $D^2$ $$0\leq x_\pm^2+y_\pm^2=a^2sin^2\theta\leq a\eqno{(44)}$$ in the equatorial planes $z_\pm=0$. It is clear that its interiors ${\dot{D}}^2$ must be identified i.e. $(x_++\epsilon,y_++\epsilon)=(x_--\epsilon,y_--\epsilon)$ as $\epsilon\to\pm 0$ for those $x_\pm$, $y_\pm$ satisfying $x_\pm^2+y_\pm^2<a^2$, and that the boundary $$x_\pm^2+y_\pm^2=a^2 \eqno{(45)}$$ is the singularity. On one ``side" of ${\dot{D}}^2$ one has the $\R^2\times S^2$ region $r^\prime>0$ (with horizons and ergosurfaces), on the other ``side" one has another $\R^2\times S^2$ region which corresponds to $r^\prime<0$, but without horizons and ergosurfaces. (In the $\R^2$ factors, one $\R$ comes from $r^\prime>0$ and $r^\prime<0$, the other from the time coordinate.) It is easily seen that in the $r^\prime\to 0_\pm$ limit, the ellipsoids (41) degenerate into the disks (44), while the hyperboloids (42) degenerate into $(z_\pm=0$ planes $\setminus {\dot{D}}^2)$. 

\

The ergosurfaces $S_\pm$ (in the $r^\prime>0$ region) are determined by the zeros of $g_{tt}$. From (31), $$r^\prime_{S_{\pm}}(\theta)=M\pm\sqrt{M^2-(a^2cos^2\theta+Q^2)}. \eqno{(46)}$$ In particular $$r^\prime_{S_{\pm}}(0)=r^\prime_{S_{\pm}}(\pi)=r^\prime_\pm \eqno{(47)}$$ i.e. $S_\pm\equiv H_\pm$ at the ``north" and ``south" poles. Replacing (46) in (36) with $F(r^\prime)=0$, we obtain for both $S_+$ and $S_-$ the surfaces of revolution $${ {{x_+}_{S_\pm}(\theta)^2+{y_+}_{S_\pm}(\theta)^2}\over{(\sqrt{r^\prime_{S_\pm}(\theta)^2+a^2})^2} }+{ {{z_+}_{S_\pm}(\theta)^2}\over{r^\prime_{S_\pm}(\theta)^2} }=1\eqno{(48)}$$ which, together with horizons $H_\pm$, ellipsoids and hyperboloids corresponding to different sets of values of $M$, $a$, and $Q^2$ are plotted in the $y_+-z_+$ plane in Figures 2, 3 and 4. For $\phi={{\pi}\over{2}}$ ($y_+-z_+$ plane or $x_+=0$), $${y_+}_{S_\pm}({{\pi}\over{2}})=\sqrt{(r^\prime_{S_\pm}({{\pi}\over{2}}))^2+a^2}=\sqrt{(M\pm\sqrt{M^2-Q^2})^2+a^2}\eqno{(49)}$$ with $r^\prime_{S_+}({{\pi}\over{2}})=M+\sqrt{M^2-Q^2}>r^\prime_+>M$ and $r^\prime_{S_-}({{\pi}\over{2}})=M-\sqrt{M^2-Q^2}<r^\prime_-<M.$ So, for $Q^2=0$, $${y_+}_{S_+}({{\pi}\over{2}})=\sqrt{4M^2+a^2}, \ \ {y_+}_{S_-}({{\pi}\over{2}})=a,\eqno{(50)}$$ and, for $Q^2>0$, $${y_+}_{S_+}({{\pi}\over{2}})=\sqrt{(M+\sqrt{M^2-Q^2})^2+a^2}, \ \ a<{y_+}_{S_-}({{\pi}\over{2}})=\sqrt{(r^\prime_{S_-}({{\pi}\over{2}}))^2+a^2}<\sqrt{{r^\prime_-}^2+a^2}<\sqrt{M^2+a^2}.\eqno{(51)}$$

\

The same structure of ellipsoids and hyperboloids are in the $y_--z_-$ plane; however, in this case, no of the ellipsoids corresponds to $H_+$ or $H_-$.

\

The spatial topology consists of two copies of $\R^3$ glued by an open disk of radious $a$ with its circular boundary being the singularity; each $\R^3$ is equivalent to $\R\times S^2$, but only one of them contains the horizons and hypersurfaces. Including the time axis, one ends with two $\R^2\times S^2$'s, an open solid cylinder $\R\times{\dot{D}}^2$, and a hollow cylinder $\R\times S^1$, the singularity. The spatial topology can be ``viewed" formally reducing the spatial dimension through the elimination of the $x$- and $x^\prime$-axis: two $\R^2$'s joined at an open segment with the singularity being its end points (Figure 5). 

\

{\bf 8. Penrose-Carter diagram}

\

For completeness, we present the Penrose-Carter diagram corresponding to the $K$ and $KN$ cases, consisting in the infinite ``vertical" repetition of the elementary cell illustrated in Figure 6. All elements: horizons and singularity are represented in the cell. The infinite tower is necessary to have a geodesically complete spacetime.

\

{\bf Acknowledgements}

\

This work was partially supported by the project PAPIIT IN105413, DGAPA, UNAM.

\

{\bf References}

\

1. Boyer, R. H. and Lindquist, R.W. {\it Maximal Analytic Extension of the Kerr Metric}. Journal of Mathematical Physics, {\bf 8} (1967), 265-281.

\

2. Carter, B. {\it Complete Analytic Extension of the Symmetry Axis of Kerr's Solution of Einstein's Equations}. Physical Review, {\bf 141} (1966), 1242-1247.

\

3. Carter, B. {\it Global Structure of the Kerr Family of Gravitational Fields}. Physical Review, {\bf 174} (1968), 1559-1571.

\

4. Drake, S. P. and Szekeres, P. {\it Uniqueness of the Newman-Janis Algorithm in Generating the Kerr-Newman Metric}. General Relativity and Gravitation, {\bf 32} (2000), 445-457.

\

5. Eddington, A. S. {\it A Comparison of Whitehead's and Einstein Formulas}. Nature, {\bf 113} (1924), 192. 

\

6. Finkelstein, D. {\it Past-Future Asymmetry of the Gravitational Field of a Point Particle}. Physical Review, {\bf 110} (1958), 965-967.

\

7. Kerr, R. P. {\it Gravitational Field of a Spinning Mass as an Example of Algebraically Special metrics}. Physical Review Letters, {\bf 11} (1963), 237-238.

\

8. Kerr, R. P. and Schild, A. {\it A New Class of Vacuum Solutions of the Einstein Field equations}. General Relativity and Gravitation, {\bf 41} (2009), 2485-2499. Original paper: R. P. Kerr and A. Schild, in: ``Atti del Convegno sulla Relativita Generale: Problemi dell' Energia e Onde Gravitazionali", G. Barb\`ere Editore, Firenze 1965, pp. 1-12.

\

9. Pleba\'nski, J. and Krasi\'nski. {\it An Introduction to General Relativity and Cosmology}, Cambridge University Press, Cambridge, (2006), p. 450.

\

10. Newman, E. T. and Janis, A. I. {\it Note on the Kerr Spinning-Particle Metric}. Journal of Mathematical Physics, {\bf 6} (1965), 915-917.

\

11. Newman, E. T., Couch, E., Chinnapared, K., Exton, A., Prakash, A., and Torrence, R. {\it Metric of a Rotating, Charged Mass}. Journal of Mathematical Physics, {\bf 6} (1965), 918-919.

\

12. Nordstrom, G. {\it On the Energy of the Gravitational Field in Einstein's Theory}. Proc. Kon. Ned. Akad. Wet., {\bf 20} (1918), 1238-1245.

\

13. Penrose, R. {\it Asymptotic Properties of Fields and Space-times}. Physical Review Letters, {\bf 10} (1963), 66-68.

\

14. Reissner, H. {\it $\ddot{U}$ber die Eigengravitation des Elektrischen Feldes nach der Einsteinschen Theorie}. Annalen der Physik, {\bf 50} (1916), 106-120. 

\

emails: brauer@ciencias.unam.mx, hugocm$_{-}$89@hotmail.com, socolovs@nucleares.unam.mx

\end

The uncharged Kerr ($K$) (Kerr, 1963) and the charged Kerr-Newman ($KN$) (Newman et al, 1965) axially symmetic stationary spacetimes have, in contradistinction to the charged static spherical Reisner-Nordstrom ($RN$) solution (Reissner, 1916; Nordstrom, 1918), an asymptotically flat region which involves, both in the Eddington-Finkelstein ($EF$) (Eddington, 1924; Finkelstein, 1958) and  Boyer-Lindquist ($BL$) (Boyer and Lindquist, 1967) coordinates, a radial cordinate taking not only positive and zero values, but also negative ones, form 0 to -$\infty$.

\centerline{\bf FIBRE BUNDLES AND GAUGE TRANSFORMATIONS}

\

{\bf 1. Principal fibre bundle $\xi$}

\

A smooth principal fibre bundle (p.f.b.) $\xi$ consists of a sextet $$\xi=(P,\pi,M,G,\xi,{\cal U})$$ where $P$ and $M$ are $m+s$ and $m$ dimensional differentiable manifolds, $G$ is an $s$ dimensional Lie group, $P\buildrel{\pi}\over\rightarrow M$ is a projection, $P\times G\buildrel{\psi}\over\rightarrow P$ is an action ($pe=p$ and $p(g_1g_2)=(pg_1)g_2$) free over $P$ ($pg=p\Rightarrow g=e$, $e$ the identity in $G$) and transitive on fibers $P_x=\pi^{-1}(\{x\})$ (if $p,q\in P$ then there exists $g\in G$ such that $q=pg$), and ${\cal U}$ is the atlas of the bundle i.e. the set of local trivializations $P_U \buildrel{\phi_U}\over\rightarrow U\times G$ with $U$ open subset of $M$, $P_U=\pi^{-1}(U)$, and $\phi_U$ a diffeomorphism $\phi(p)=(\pi(p),\gamma_U(p))$ with $\gamma_U(pg)=\gamma_U(p)g$. 

\

$\xi$ is denoted by $G^s\to P^{m+s}\buildrel{\pi}\over\rightarrow M^m$.

\

A section of $P$ is a smooth function $M\buildrel{\sigma}\over\rightarrow P$ such that $\pi\circ\sigma=Id_M$.

\

{\it 1.1 Gauge group of} $\xi$: ${\cal G}(\xi)$

\

${\cal G}(\xi)$ is the set of diffeomorphisms $P\buildrel{\alpha}\over\rightarrow P$ such that the following diagram commutes: $$\matrix{P\times G & \buildrel{\alpha\times Id_G}\over\rightarrow & P\times G \cr \psi\downarrow & & \downarrow\psi \cr P & \buildrel{\alpha}\over\rightarrow & P\cr \pi\downarrow & & \downarrow\pi\cr M & \buildrel{Id_M}\over\rightarrow & M\cr}$$ i.e. $\alpha(pg)=\alpha(p)g$ and $\pi(\alpha(p))=\pi(p)$ (so $\alpha(p)\in P_{\pi(p)}$). ${\cal G}(\xi)$ is called the {\it group of gauge transformations} or {\it vertical automorphisms} of $\xi$.

\

{\it 1.2 Local form of} ${\cal G}(\xi)$

\

{\it Proposition}: For each $(U,\phi_U)\in{\cal U}$ there exists a smooth function $$\alpha_U:U\to G, \ x\mapsto\alpha_U(x)$$ which determines $\alpha\in{\cal G}(\xi)$ for all $p$ in $P_U$.

\

{\it Proof}: $\alpha(\sigma_U(x))=\sigma_U(x)\alpha_U(x)$ for some unique $\alpha_U(x)\in G$, where $U\buildrel{\sigma_U}\over\rightarrow P_U$ is the local section of $P$ given by $\sigma_U(x)=\phi_U^{-1}(x,e)\in P_U$. If $p\in P_x$, then $p=\sigma_U(x)h$ for some unique $h\in G$; so $\alpha(p)=\alpha(\sigma_U(x)h)=\alpha(\sigma_U(x))h=(\sigma_U(x)\alpha_U(x))h=\sigma_U(x)(\alpha_U(x)h)$ i.e. $\alpha_U$ gives $\alpha$ for all $p\in P_x$ for all $x\in U$. qed

\

{\it Corollary}: If $\xi$ is trivial i.e. if $P\cong M\times G$, then $U=M$ and there exists $\alpha_M:M\to G$. qed

\

Clearly, if $\xi$ is trivial, then $M\buildrel{\sigma}\over\rightarrow P$ with $\sigma(x)=\phi^{-1}(x,e)$ is a section of $P$, where $\phi$ is the diffeomorphism $P\buildrel{\phi}\over\rightarrow M\times G$. The other way around: if $\xi$ has a section $M\buildrel{\sigma}\over\rightarrow P$, $\sigma(x)=p$, then $P\buildrel{\phi}\over\rightarrow M\times G$ with $\phi(q)=(\pi(q),g), \ q=pg$, is a trivialization of $P$. So, a p.f.b. is trivial if and only if it has a (global) section.  

\

{\it 1.3 Example: Frame bundle of a differentiable manifold}

\

$$\xi_{\cal F}: \ \ (GL_m(\R))^{m^2}\to (FM^m)^{m^2+m}\buildrel{\pi_{\cal F}}\over\rightarrow M^m$$ where $$FM^m=\bigcup_{x\in M^m}\{x\}\times\{(v_{1x},\cdots,v_{mx})\},$$ with $$(v_{1x},\cdots,v_{mx})\equiv r_x$$ an ordered basis ({\it Vielbeine}) of $T_xM^m$, $FM^m\times GL_m(\R)\buildrel{\psi}\over\rightarrow FM^m, \ \psi(r_x,a)=(v_{\nu x}a^\nu_1,\cdots,v_{\nu x}a^\nu_m)$, local coordinates on $FM^m$ given by $x^\rho(x,r_x)=x^\rho(x)$ and $X^\mu_\nu(x,r_x)=v^\mu_{\nu x}$, and $\pi_{\cal F}(x,r_x)=x$. If $\{{{\partial}\over{\partial x^\mu}}\}_{\mu=1}^m$ is a local coordinate basis of $\Gamma(TU)$, then $v_{kx}=v_ k^\mu(x){{\partial}\over{\partial x^\mu}}\vert_x$, with $x\in U$, $k=1,\dots,m$.

\

In $\xi_{\cal F}$ there exists the {\it canonical form} $\theta:FM^m\to T^*FM^m\otimes\R^m$, $\theta((x,r_x))=((x,r_x),\theta_{(x,r_x)})$, and $\theta_{(x,r_x)}:T_{(x,r_x)}FM^m\to\R^m$ given by the composition $\tilde{r}_x\circ d\pi_{\cal F}\vert_{(x,r_x)}$ where $\tilde{r}_x:T_xM^m\to \R^m$, $\tilde{r}_x(\lambda^\nu v_{\nu x})=(\lambda^1,\dots,\lambda^m)$. So, $\theta$ solders the tangent spaces of $M^m$ to the tangent spaces of $FM^m$. 

\

{\bf 2. Associated bundle} $\xi_F$

Let $\xi$ be a p.f.b., $F$ an $n$ dimensional differentiable manifold, and $G\times F\buildrel{\mu}\over\rightarrow F$ a left action of $G$ on $F$. Then $\xi$ and $\mu$ induce the bundle $$\xi_F=(P_F,\pi_F,M,F,\mu,{\cal U}_F)$$ with total space the $n+m$ dimensional differentiable manifold $$P_F=P\times_GF=\{[(p,f)]\}_{(p,f)\in P\times F}, \ [(p,f)]=\{(pg,g^{-1}f)\}_{g\in G},$$ fiber $F$, projection $$P_F\buildrel{\pi_F}\over\rightarrow M, \ \pi_F([(p,f)])=\pi(p),$$ and atlas ${\cal U}_F$ given by the local trivializations $$P^F_U\buildrel{\phi^F_U}\over\rightarrow U\times F, \ \phi^F_U([(p,f)])=(\pi(p),\gamma_U(p)f),$$ where $P^F_U=\pi_F^{-1}(U)$. 

\

($\phi^F_U$ is well defined since $\phi^F_U([(pg,g^{-1}f)]=(\pi(pg),\gamma_U(pg)g^{-1}f)=(\pi(p),\gamma_U(p)gg^{-1}f)=\phi^F_U([(p,f)]$.)

\

$\xi_F$ is called an {\it associated bundle} (a.b.) and is denoted by $F^n-P_F^{n+m}\buildrel{\pi_F}\over\rightarrow M^m$.

\

The space of {\it sections} of $\xi_F$ is the set of smooth functions $$\Gamma(P_F)=\{s:M\to P_F \ \vert \ \pi_F\circ s=Id_M\}.$$

\

The space of {\it equivariant functions} $P-F$ is the set of smooth functions $$\Gamma_{eq.}(P,F)=\{\gamma:P\to F \ \vert \ \gamma(pg)=g^{-1}\gamma(p)\}.$$

\

{\it Proposition}: There is a one-to-one correspondence between $\Gamma(P_F)$ and $\Gamma_{eq.}(P,F)$.

\

{\it Proof}: Given $s\in \Gamma(P_F)$, there exists $\gamma_s:P\to F$, $\gamma_s(p)=f$ where $s(\pi(p))=[(p,f)]$. Given $\gamma$, there exists $s_\gamma:M\to P_F$, $s_\gamma(x)=[(p,f)]$ with $\gamma(p)=f$ and $p\in\pi^{-1}(\{x\})$. qed

\

This is summarized in the following diagram:

\

$$\matrix{G & & F\cr \downarrow & \gamma\nearrow & \vert\cr P & & P_F\cr \pi\downarrow & & s\uparrow\downarrow\pi_F\cr M & & M\cr}$$

\

If $F$ is a vector space $V$ and $\mu$ is a linear representation of $G$ over $V$, then $\xi_V$ is a {\it vector bundle} (v.b.) associated to $\xi$. Typically $V=\R^n$ or $\C^n$. If $\{e_1,\cdots,e_n\}$ is a basis of $V$, then $$U\buildrel{\sigma^V_{U,k}}\over\rightarrow P^V_U, \ \sigma^V_{U,k}(x)=[(p,(\gamma_U(p))^{-1}e_k)], \ p\in P_x,$$ $k=1,\cdots,n$, is a set of $n$ linearly independent local sections of $\xi_V$. Typically, $V$ is the representation space of an irreducible representation of $G$.

\

{\it 2.1 Gauge transformations of sections}

\

Let $\alpha\in{\cal G(\xi)}$ and $s\in\Gamma(P_F)$, then $\alpha^*(\gamma_s)=\gamma_s\circ\alpha\in\Gamma_{eq.}(P,F)$ since $\gamma_s\circ\alpha(pg)=\gamma_s(\alpha(pg))=\gamma_s(\alpha(p)g)=g^{-1}\gamma_s(\alpha(p))=g^{-1}\gamma_s\circ\alpha(p)=g^{-1}\alpha^*(\gamma_s)(p)$. Then $$\tilde{\alpha}(s)=s_{\alpha^*(\gamma_s)}$$ is the gauge transformed of the section $s$, with $$\tilde{\alpha}(s)(x)=[(p,g^{-1}f)]$$ where $s(x)=[(p,f)]$, $\alpha(p)=pg$, and $p\in\pi^{-1}(\{x\})$.

\

{\it 2.2 Example: Bundle of tensors $T^r_sM^m$ over a differentiable manifold}

\

$G=GL_m(\R)$, $F=V=\R^{m^{r+s}}$, $r,s\in\{0,1,2,3,\dots\}$, $$\mu(a,\vec{\lambda})_{j_1,\dots,j_s}^{i_1,\dots,i_r}=a^{i_1}_{k_1}\dots a^{i_r}_{k_r}(a^{-1})^{l_1}_{j_1}\dots (a^{-1})^{l_s}_{j_s}\lambda^{k_1\dots k_r}_{l_1\dots l_s}.$$ One has the bundle isomorphism $\varphi$ given by the diagram: 

$$\matrix{\R^{m^{r+s}} & & \R^{m^{r+s}} \cr \vert & & \vert \cr FM^m\times_{GL_m(\R)}\R^{m^{r+s}} & \buildrel{\varphi}\over\longrightarrow & T^r_sM^m \cr \pi_V\downarrow & &  \downarrow\pi^r_s \cr M^m & \buildrel{Id_{M^m}}\over\longrightarrow & M^m \cr},$$ where
$\varphi([(r_x,\vec{\lambda})])=\sum^m_{i_k,j_l=1}\lambda^{i_1\dots i_r}_{j_1\dots j_s}v_{i_1x}\otimes\dots v_{i_rx}\otimes w_x^{j_1}
\otimes\dots\otimes w_x^{j_s}$, with $\{w_x^i\}$ the dual basis of $\{v_{jx}\}$: $w_x^i(v_{jx})=\delta^i_j.$

\

{\bf 3. Connection in a principal fibre bundle}

\

{\it 3.1 Gauge transformations of connections}

\

{\bf 4. Covariant derivative of sections}

\

{\it 4.1 Gauge transformations of covariant derivative}

\

{\it 4.2 Local form of covariant derivative}

\

{\it 4.3 Parallel transport}

\

\

\

\

\

\

\

\

\

e-mail: socolovs@nucleares.unam.mx

\end